\documentclass[twocolumn,showpacs,preprintnumbers,amsmath,amssymb]{revtex4}

\usepackage{graphicx}
\usepackage{amsmath}
\usepackage{bm}

\begin{document}

%\preprint{}

\title{
Shape of Heteroepitaxial Island Determined by Asymmetric Detachment 
}
\author{Yukio Saito}  \email{yukio@rk.phys.keio.ac.jp}
\author{Ryo Kawasaki} 

\affiliation{
Department of Physics, Keio University, Yokohama 223-8522, Japan
}

\begin{abstract}
Square lattice gas models for heteroepitaxial growth are studied by
means of kinetic Monte Carlo simulations,
in order to find a possible origin of anisotropic island shape
observed in growth experiments of
long organic molecules.
When deposited molecules form clusters irreversibly at their encounter
during surface diffusion, 
islands grow in a ramified dendritic shape, similar to DLA.
Introduction of molecular detachment from edges makes islands compact
with smooth edges.
Tilting of adsorbed long molecules or steps in a vicinal substrate 
may induce orientation-dependence in the detachment rate of
edge molecules from an island.
In simulations with orientation-dependent detachment rates, a clear 
anisotropy in an island shape is observed.
Shape anisotropy on a vicinal substrate is enhanced as steps 
get dense, in agreement to the experimental observation.\\
\end{abstract}


\sloppy

\maketitle

\section{Introduction}

There have been increasing interests in heteroepitaxial growth 
of organic semiconductors
in relation to the fabrication of thin and deformable
microelectronic or optoelectronic devices
\cite{tang+87,nelson+98,dimitrakopoulos+99}.
To accomplish a high standard in electrical properties,  
improvement of the film quality is indispensable,
and fundamental researches on growth mechanisms of 
adsorbed organic molecules are undertaken
\cite{heringdorf+01,nishikata+07}.
In an experiment of pentacene (Pn,  C$_{14}$H$_{22}$) growth on Si(001)
\cite{heringdorf+01},
authors concluded that an organic thin-film growth is similar
to the epitaxial growth of inorganic materials, 
such as the formation of fractal ramified islands
like diffusion-limited aggregates (DLA)
\cite{witten+81,witten+83}.
On the other hand, 
in an experiment of Pn growth on a hydrogen-terminated Si(111) surface
[H-Si(111)]
\cite{nishikata+07},
a peculiarity is found.
On a flat H-Si(111) surface islands are compact with smooth edges and
isotropic, whereas on a vicinal surface
islands are dendritic in the step-down orientation while 
edges in the step-up orientation remain compact:
Each island has a strong anisotropy in its shape on a vicinal substrate.
Island shape anisotropy, however,  is not specific to a vicinal surface.
It is observed in the vapor growth of another organic molecule, 
a behenic acid, even on a flat singular substrate, namely on an oxidized GaAs
\cite{kambe+07}.
In the present paper we analyze some model systems 
which lead to shape anisotropy theoretically.

In the heteroepitaxial growth of inorganic materials,
island shape is mainly controlled by the competition of 
two processes after an adatom deposition; 
an adatom diffusion on a terrace surface and 
an incorporation kinetics at island edges
\cite{barabasi+95,pimpinelli+98,venables00,michely+03}.
With a fast incorporation at island edges,
diffusion controls the growth and edges undergo a morphological 
instability to form dendrites.
In the extreme case of irreversible solidification
such that an adatom once solidified never detaches again from the island edge,
then an island takes a ramified irregular shape with many branches
\cite{hwang+91}.
Its structure is similar to the DLA without a characteristic length, 
and is called a fractal.
An island is isotropic or symmetric by reflecting the lattice symmetry.

To provide a compact island shape, edge smoothening processes are
necessary, such as an edge diffusion or a detachment from the edge.
If the detachment rate is high, adatoms loosely attached to the edge will
readily detach and they are hardly incorporated in two-dimensional (2D) 
islands.
Then, the slow incorporation kinetics governs the growth to
make islands compact with smooth edges
\cite{bartelt+94,bales+95,saito03}.
Islands may be round at high temperatures or polygonal
at low temperatures.

In the heteroepitaxial growth of organic molecules,
one has to take an additional feature into account, 
i.e. the size of an adsorbed organic molecule. 
It can no longer be regarded as a point object
as in inorganic cases so far discussed.
Both pentacene and behenic acid molecules are flat and elongated in shape.
When they are adsorbed on a substrate,
molecules can be normal or lateral to the substrate surface
\cite{takiguchi+95}.
Pn molecules are known to grow a wetting film with the molecular long axis 
normal to the H-Si(111) surface, namely in a standing-up orientation
\cite{shimada+05}. 
Behenic acid is also known to be in a stand-up orientation but 
with a small tilting on a substrate.
We assume that this molecular orientation affects
the detachment rate of molecules from island edges, and
leads to anisotropy in an island shape.

In a previous study\cite{saito+07} we proposed a simple lattice gas model,
and studied general aspects of the effect of anisotropic detachment 
on an island shape.
In the next \S 2 we introduce the model with simulation results of 
island shapes on a flat substrate.
The model is extended  in \S 3 so as to include the effect of steps explicitly 
for the case of adsorption on a vicinal substrate.
The last \S 4 summarizes the result.

\section{2D Islands on a Flat Substrate}
When molecules in a 2D crystal
stack almost normal to a substrate surface but with a finite tilting,
as to the $a$-axis in the case of a behenic acid
\cite{kambe+07},
kinetics at island edges may depend on 
relative orientations of the edge and of the molecular tilting.
To an edge where the molecular inclination restricts the incorporation
space, diffusing molecules may be difficult to be attached to the edge,
but once attached they may be difficult to be detached from this edge. 
Depression of the attachment rate slows down the incorporation kinetics, 
and the edge becomes smooth. 
On the contrary, suppression of detachment increases incorporation
and leads to a diffusional instability of a smooth edge.
So far the relation between the molecular tilting and the orientation of
dendritic edge is not identified experimentally
\cite{kambe+07},
we here assume 
simply that a molecular tilting affects
detachment process such that the detachment rate of a molecule depends on
the edge orientation.
In our theoretical treatment, we forget about the finite size 
and shape of an adsorbing organic molecule, but consider it 
implicitly in the orientation-dependence of the detachment rate.

We perform kinetic Monte Carlo simulations
\cite{saito+07,bortz+75}
of a lattice gas models of point molecules depositing on a square
substrate with a deposition rate $F$ per area.
Adsorbed molecules diffuse on a surface with a diffusion constant $D_s$.
When two diffusing molecules meet, they make a bond to lower an energy
and form a cluster.
Clusters enlarge their sizes by incorporating diffusing molecules
at edges. On the other hand,
molecules at edges can detach from the islands
when they are loosely attached: 
Those particles with
a single bond to the island can break the bond to migrate out
on the substrate surface with a rate 
$D_e$, but molecules with more than two bonds are assumed immobile.
When a particle is deposited above an island, it diffuses on
the island terrace till it reaches an edge, and steps it down
to be incorporated into the island.
Evaporation of adsorbed molecules from the surface is excluded,
since the growth is assumed to take place at a low temperature.
When there is no detachment, grown islands take a ramified dendritic form
with irregular fine branches, similar to the DLA
\cite{witten+81,witten+83}.
With a small detachment, one obtains dendritic aggregates
with fat branches, as shown in Fig. 1(a). Here, parameters are
$D_s/F=10^{12}$ and $D_e/F=10^4$ with a system size $L^2=500^2$
and the coverage $\Theta=0.1$. 
Although islands interfere with each other in a diffusion field,
they have almost a symmetric shape.
By applying the box-counting method, one obtains a rough estimate of
the fractal dimension as $d_f \approx 1.70$.

%%%%%%%%%%%%%%%%%%%%%
\begin{figure}[h]
\begin{center} 
\includegraphics[width=0.32\linewidth]{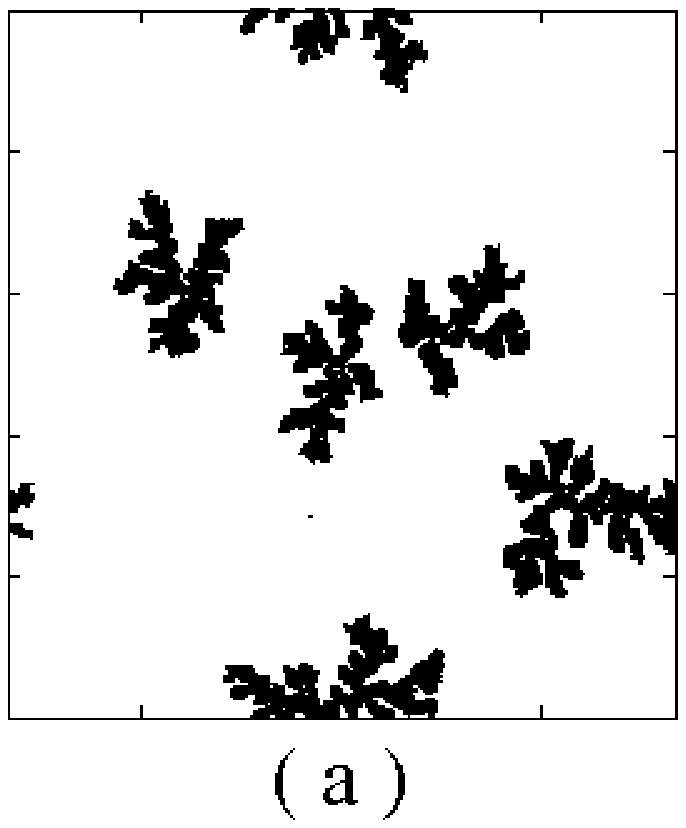}
\includegraphics[width=0.32\linewidth]{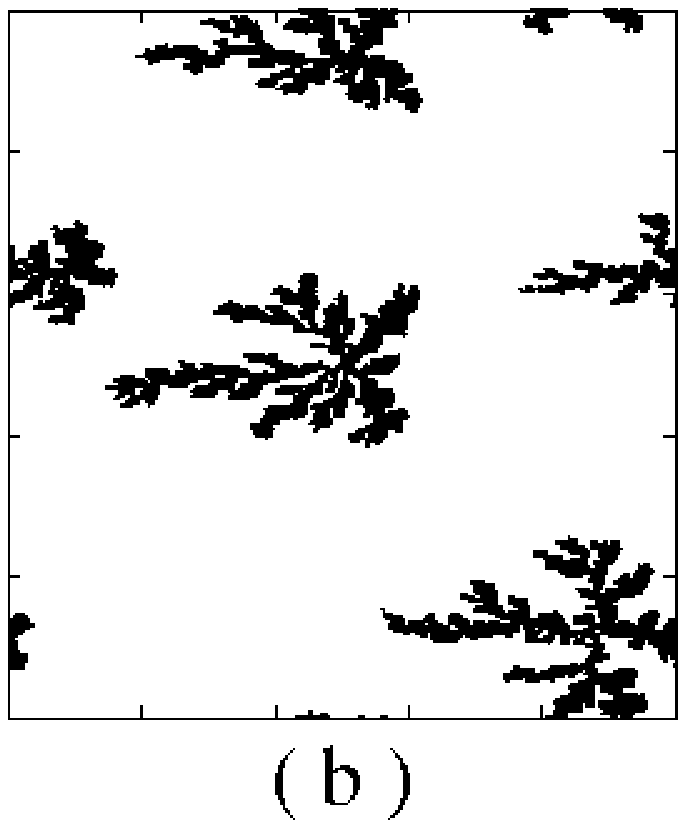}
\includegraphics[width=0.32\linewidth]{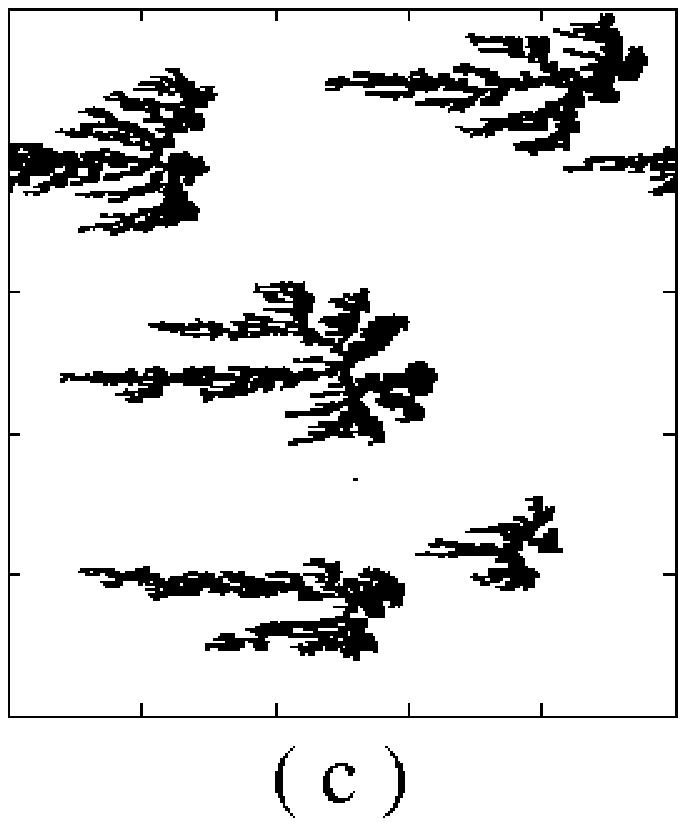}\\
\end{center} 
\caption{Islands on a substrate with a size $500^2$
with a surface diffusion $D_s/F=10^{12}$ at a coverage $\Theta=0.1$.
A detachment rate is $D_e/F=10^4$, and
(a) isotropic, and 
(b) and (c) anisotropic.
Detachment from the left edge is suppressed by a factor (b) $R_L=0.1$, and  
(c) $R_L=0.01$.
}
\label{fig1}
\end{figure}
%%%%%%%%%%%%%%%%%%%%%

We now introduce an anisotropy in the rate of detachment from edges.
 On the left edge of an island,
for instance, we arbitrary reduce the rate of detachment by $R_L$.
Island morphology simulated by means of kinetic Monte Carlo simulation
\cite{bortz+75}
becomes anisotropic, as shown in Fig. 1(b) and 1(c).
When the detachment from the left edge is reduced by a factor 10 as $R_L=0.1$
(Fig. 1(b)),
the left edges of islands grow faster than those in the other orientations,
and the primary branch to the left extends about three times more than 
to the right. 
The fractal dimension by box counting remains the same value $d_f\approx 1.7$.
On further reducing the detachment rate from the left edges to $R_L=0.01$
(Fig. 1(c)),
islands point sharply to the left with the left-to-right ratio
ranging about 3 to 6, depending on environment.
Primary branches extending to the left are fine and have less side-branches,
compared to Fig. 1(a) and 1(b), and primary branches to other orientations
have many secondary branches extending long to the left.
Pointing structure leads to the small fractal dimension of $d_f \approx 1.6$
by the box-counting method.
Island morphology in Fig. 1(c)
looks quite similar to those observed for
the behenic acid islands on an oxidized GaAs
\cite{kambe+07}.

Instead of suppressing detachment from the left edge, the enhanced detachment
from the righ edge leads to anisotropic island shape \cite{saito+07}.
In this case, however, islands extend vertically in $y$ direction, and 
the obtained shape looks different to what is observed in the experiment
 \cite{kambe+07}.
On increasing the detachment rate $D_e/F$, islands become more compact with 
smooth edges as a skeletal shape or even to a complete square.
The effect of anisotropic detachment in these cases have been discussed 
previously
\cite{saito+07}.

\section{Islands on a Vicinal Surface}
We know consider the case of a heteroepitaxial growth on a vicinal surface,
which might be relevant to Pn growth on H-Si(111)
\cite{nishikata+07}.
On a flat terrace, 
Pn molecules form 2D islands, and they
 are compact with an isotropic shape.
Therefore, Pn molecules might often detach from island edges.
On a vicinal H-Si(111) surface, 
Pn islands show different morphology in the upward and downward directions
relative to the substrate steps.
Island edges extending to the step-down direction are pointed
in a dendritic form,
while those  to the step-up directions remain to be smooth and round.

%%%%%%%%%%%%%%%%%%%%%
\begin{figure}[h]
\begin{center} 
\includegraphics[width=0.6\linewidth]{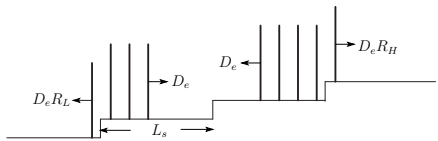}
\end{center} 
\caption{
Detachment rates of adsorbed long molecule from islands on a vicinal surface.
}
\label{fig2}
\end{figure}
%%%%%%%%%%%%%%%%%%%%%

Since Pn is known to grow in a standing-up orientation
\cite{shimada+05},
an interaction between Pn molecules is stronger than that 
between Pn and the substrate.
Let us now
imagine that a Pn crystalline film is growing on a vicinal surface
with a positive slope, as shown in Fig. 2. When a film grows to the left
 in a step-down direction and
reaches an upper side of the descending step, as a left island in Fig. 2,
further crystal growth takes place by incorporating
a Pn molecule from the lower terrace. In this case, the whole length of a
newly attached molecule interacts with
the island edge and the step ledge, and it is hard to be detached back to the
lower terrace.
Also, the free upper part of the long molecule may in general 
be susceptible to thermal fluctuation, and there may be a substantial
entropy contribution which reduces the bonding free energy.
Attachment from the lower terrace may lessen this entropy contribution
and enhances the Pn-Pn bonding.
Thus, the detachment rate from the lower terrace is diminished 
from the normal rate $D_e$ on a flat terrace
by a factor $R_L<1$ as to $D_eR_L$.
On the contrary, when the island grows to the right in the step-up direction
and reaches the lower side of the ascending step,
as a right island in Fig. 2, 
the next Pn molecule to be incorporated is located on the higher terrace,
and its upper part does not contribute to the Pn-Pn bonding. 
Since the substrate step has a height of about one fourth of the length of 
a Pn molecule, 
 a loss in the bonding energy may be significant.
Accordingly, those molecules incorporated into an island edge
from the higher terrace are easily detached back to the higher terrace with
an enhancement factor $R_H>1$.
We now study the effect of this anisotropic detachment 
with rate modification factors $R_H$ and $R_L$
on the island morphology.

In kinetic Monte Carlo simulation of heteroepitaxial growth
on a vicinal surface, 
steps are separated with a distance $L_s$.
After a deposition, molecules perform 
$D_s/F$ times of diffusive migrations until a monolayer is covered, 
but during this period molecules collide to form clusters.
Steps are assumed to have no effect on the surface diffusion, for simplicity.
A large value of surface diffusion constant $D_s/F=10^{15}$
is chosen in order to grow
 only a single island in our system of a size $L^2=500^2$.
From edges of an island on a flat terrace, molecules with a single bond
detach with a rate $D_e/F=10^{5}$.
Above a descending step, detachment is enhanced by a factor $R_H=10$,
but in front of the step detachment is assumed completely forbidden, $R_L=0$,
corresponding to an extreme case.

%%%%%%%%%%%%%%%%%%%%%
\begin{figure}[h]
\begin{center} 
\includegraphics[width=0.24\linewidth]{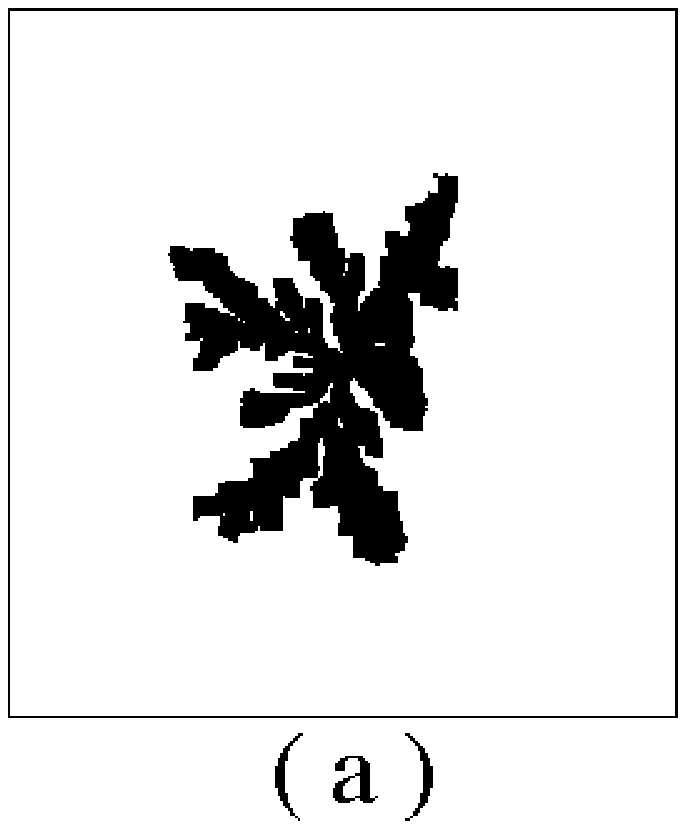}
\includegraphics[width=0.24\linewidth]{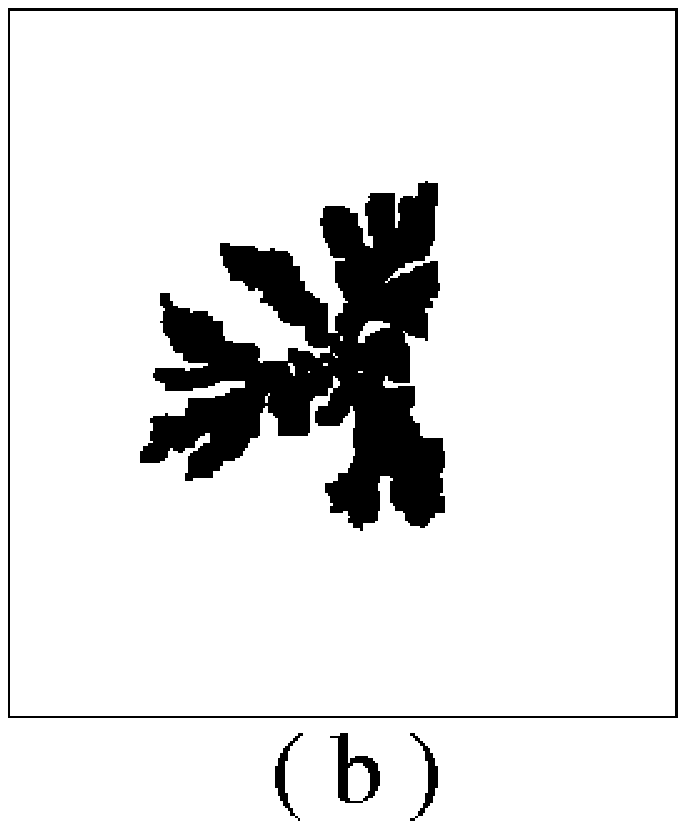}
\includegraphics[width=0.24\linewidth]{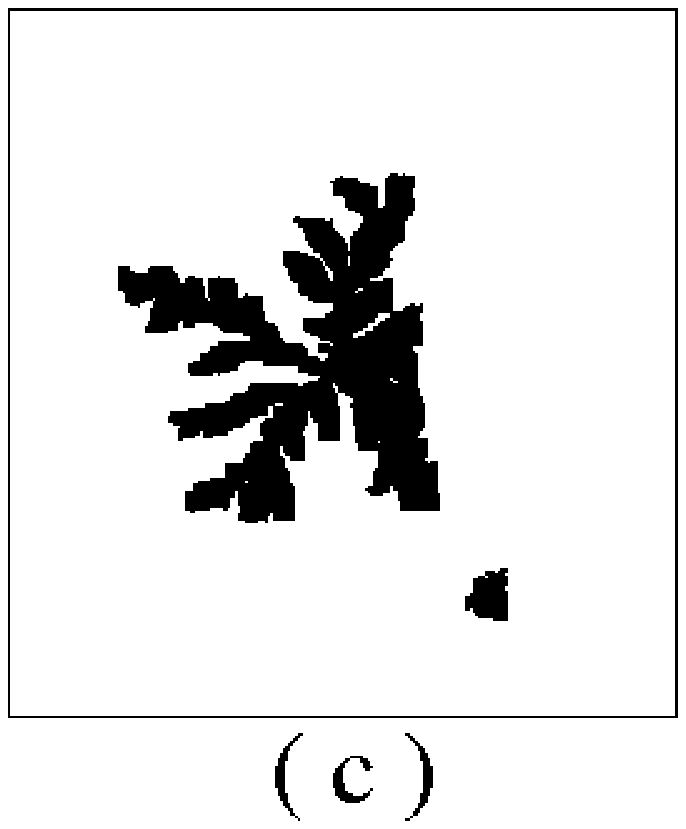}
\includegraphics[width=0.24\linewidth]{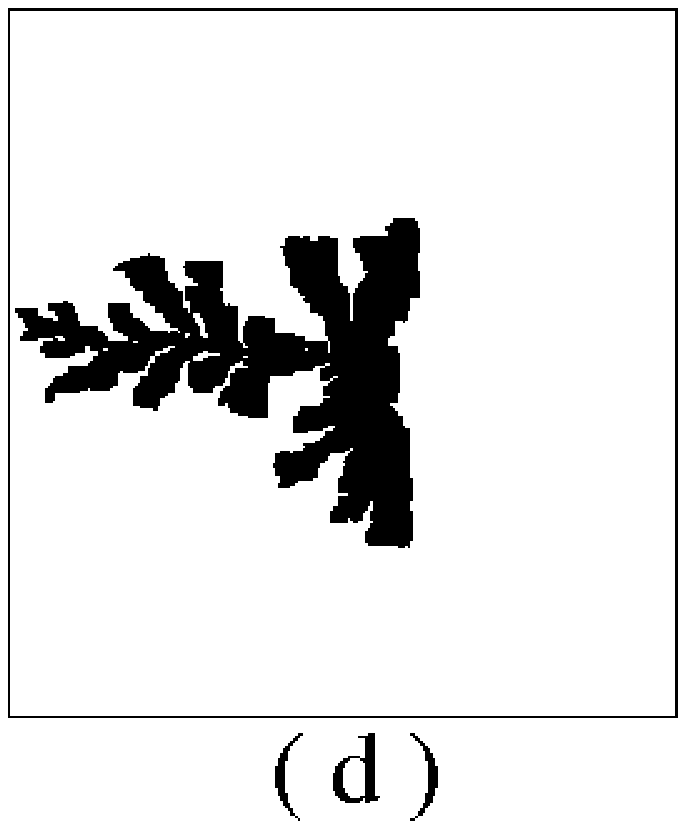}
\end{center} 
\caption{An island on a vicinal substrate surface.
The surface diffusion is $D_s/F=10^{15}$ and the detachment rate 
on a flat terrace is $D_e/F=10^5$.
Detachment to the higher terrace (to the right) is enhanced  
by a factor $R_H=10$ and 
to the lower terrace (to the left) is completely suppressed: $R_L=0$.
Step separations are $L_s=$(a) 5, (b) 4, (c) 3 and (d) 2.
The coverage is $\Theta=0.1$ on a substrate with a size $500^2$.
}
\label{fig3}
\end{figure}
%%%%%%%%%%%%%%%%%%%%%

By initially providing an embryo of a size 2 by 2 at the center of the system,
an island is nucleated and grows to the coverage $\Theta=0.1$.
As the step separation decreases from $L_s=5$ to 2,
the island shape alters as shown in Fig. 3(a) to 3(d).
Even with a very large surface diffusion, a second island is nucleated
accidentally in Fig. 3(c).
At a large step separations $L_s=5$ and 4, 
the island has only a weak left-right anisotropy,
as in Fig. 3(a) and 3(b). 
When steps come closer as $L_s=3$ and 2, anisotropy becomes obvious,
as in Fig. 3(c) and 3(d).

%%%%%%%%%%%%%%%%%%%%%
\begin{figure}[ht]
\begin{center} 
\includegraphics[width=0.5\linewidth]{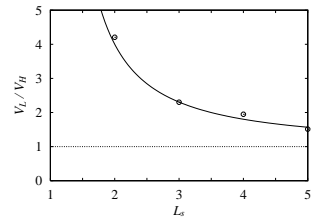}
\end{center} 
\caption{Ratio of the edge velocities to the lower side $V_L$ to 
the higher side $V_H$ as a function of the step separation $L_s$.
A curve is a guide for eyes.
The surface diffusion is $D_s/F=10^{15}$ and the detachment rate is 
$D_e/F=10^5$ with an enhancement factor to the uphill $R_H=10$ and a complete 
suppression to the downhill $R_L=0$.
}
\label{fig4}
\end{figure}
%%%%%%%%%%%%%%%%%%%%%

In order to quantify the shape anisotropy, we measure the ratio $V_L/V_H$
of the front velocities 
 to the lower side $V_L$ and to the higher side $V_H$
as a function of the step separation. The result shown in Fig. 4
clearly demonstrates that an anisotropy increases as the
step separation decreases, in a qualitative agreement to the experimental
result obtained in Pn system
\cite{nishikata+07}.

%%%%%%%%%%%%%%%%%%%%%
\begin{figure}[h]
\begin{center} 
\includegraphics[width=0.32\linewidth]{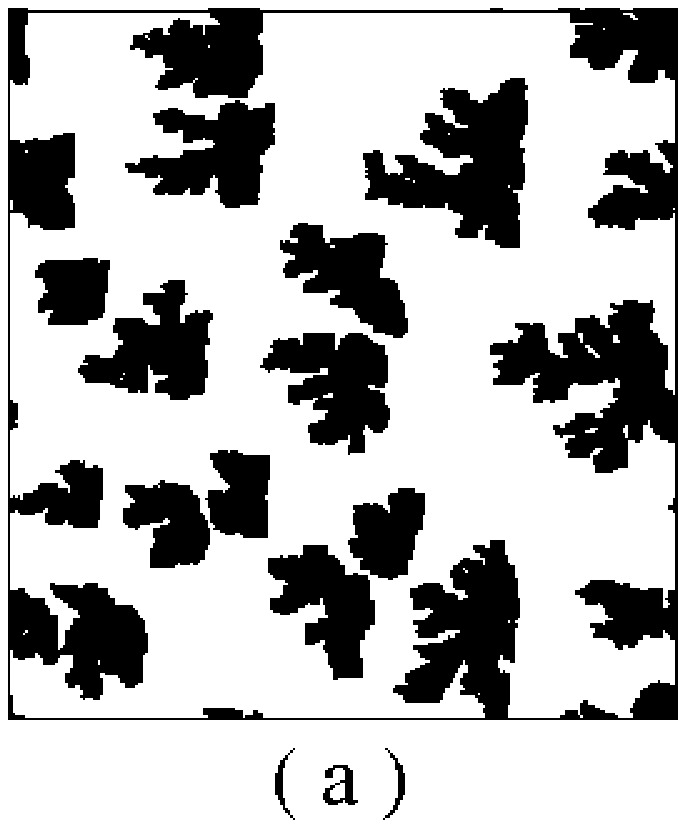}
\includegraphics[width=0.32\linewidth]{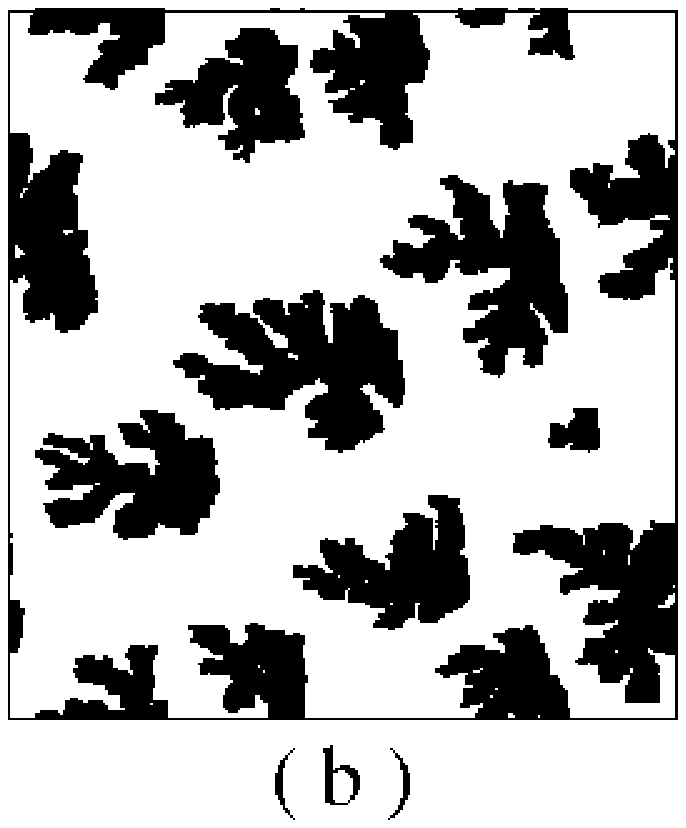}
\includegraphics[width=0.32\linewidth]{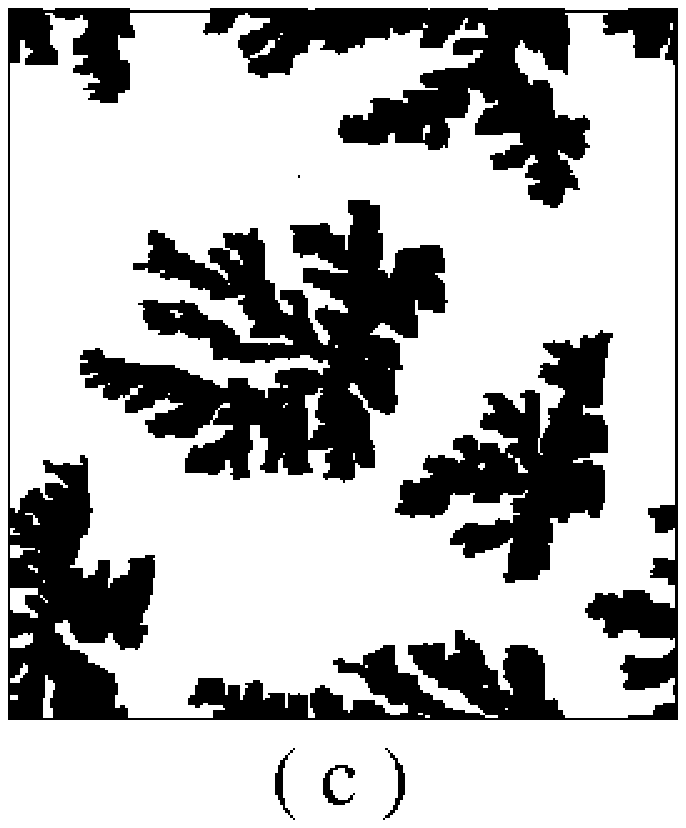}
\end{center} 
\caption{Island distribution on a vicinal surface with a size $500^2$ 
at a coverage $\Theta=0.3$.
Step separation is
$L_s=2$ and surface diffusion constants are
(a) $D_s/F=10^{10}$, (b) $10^{11}$, and (c) $10^{12}$. 
Other parameters are $ D_e/F=10^4,~R_H=5, R_L=0$.
}
\label{fig5}
\end{figure}
%%%%%%%%%%%%%%%%%%%%%

So far, we discussed the shape of a single island.
In a large area of the vicinal surface many islands are nucleated 
and they affect their mutual growth and shape through the diffusion field.
In the Pn experiment one may identify some order in spatial arrangement
of islands
\cite{nishikata+07}:
Islands look to align in queues parallel to the tilting direction 
of a vicinal substrate, with a dendritic left edge of one island 
being contiguous to the smooth right edge of a left one.
( See, for example, Fig.2 (c) and (d) in Ref.\cite{nishikata+07}. )

In our system with a limited size $L^2=500^2$, island density is controlled
by varying  the surface diffusion constant $D_s$.
In some combinations of parameter values, 
simulation can produce island arrangements 
with some spatial ordering, as shown in Fig. 5.
The height of the vicinal surface is increasing to the right with 
ascending steps at the separation $L_s=2$.
Molecules are deposited to the coverage $\Theta=0.3$.
The surface diffusion constant is set at (a) $D_s/F=10^{10}$, 
(b) $10^{11}$, and (c) $10^{12}$, by keeping the detachment rate fixed to
$D_e/F=10^{4}$, and the enhancement 
factor $R_H=5$ and no detachment to the step-down direction $R_L=0$.
At the smallest surface diffusion $D_s/F$ in Fig. 5(a), there are many
islands distributed almost randomly.
The stochastic noise associated to random deposition
is frozen in during the nucleation process.
In Fig. 5(b), as the surface diffusion increases, 
island density decreases and the diffusion process seems 
to average out the deposition noise and to
induce some correlation in islands arrangement. 
Islands do not seem to align perpendicular to the step orientation.
One might notify similar tilting in islands arrangement
in the Pn experiment
\cite{nishikata+07}.
For larger $D_s$ in Fig. 5(c),
island density becomes so low that it is difficult to identify spatial correlation
for certain.

%%%%%%%%%%%%%%%%%%%%%
\begin{figure}[h]
\begin{center} 
\includegraphics[width=0.32\linewidth]{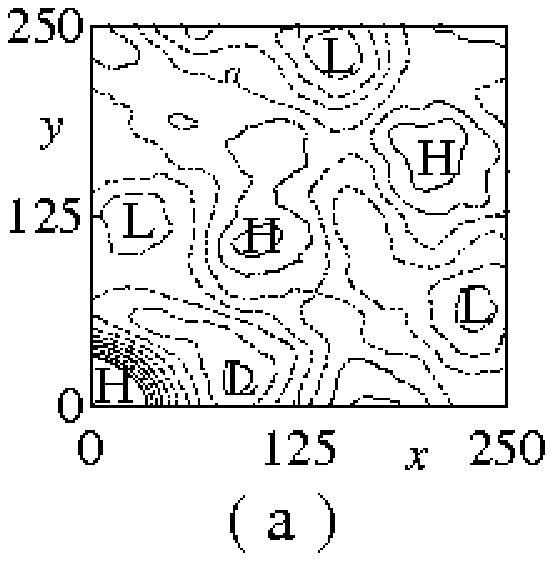}
\includegraphics[width=0.32\linewidth]{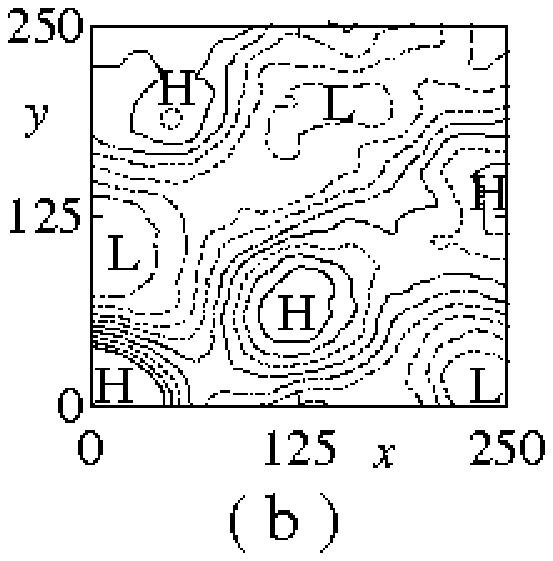}
\includegraphics[width=0.32\linewidth]{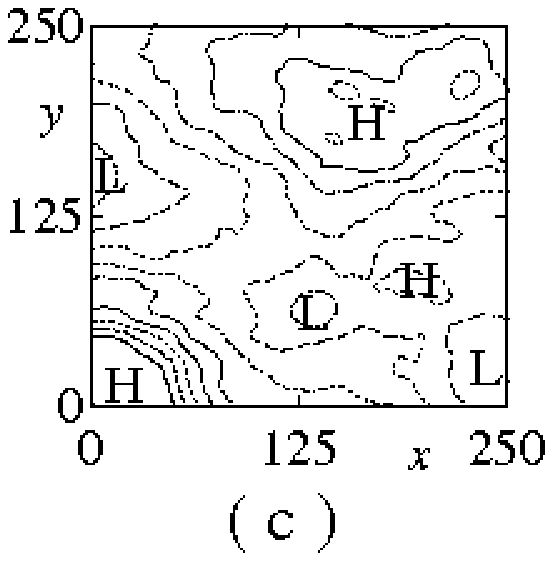}
\end{center} 
\caption{Spatial correlation function $g(\bm r)$ corresponding to
Fig. 5. "H" marks the high hill-top, and "L" marks the low valley.
}
\label{fig6}
\end{figure}
%%%%%%%%%%%%%%%%%%%%%

%%%%new
In order to evaluate spatial correlation explicitly, 
we calculated the correlation function
\begin{align}
g(\bm r)=\langle \sum_i n(\bm r+\bm r_i) n(\bm r_i) \rangle /N
\label{eq1}
\end{align}
where the occupation variable $n(\bm r_i)$ is unity when the lattice site $i$
is occupied by an admolecule, and zero when the site $i$ is empty.
Since the system is periodic both in $x$ and $y$ directions with a 
periodicity $L$, the correlation function has the 
symmetry $g(x,y)=g(L-x,y)=g(x,L-y)=g(L-x,L-y)$. Thus, 
we obtained the height contour of $g(x,y)$ for $0 \le x,y \le L/2$, 
as shown in Fig. 6.
A mark "H" represents the high hill-top with a strong correlation, 
and a mark "L" represents the low valley with a weak correlation.
From Fig.6(b) one clearly observes that the pattern has a strong periodicity
with a period $\bm p=(L/4, L/8)$ at a surface diffusion $D_s/F=10^{11}$.
With a smaller diffusion $D_s/F=10^{10}$, average island density gets higher 
as in Fig. 5(a), and the periodicity is not commensurate to the system size. 
Therefore, $g(\bm r)$ has a rather broad first peak around $\bm r=(100, 100)$
in Fig. 6(a).
With a larger diffusion $D_s/F=10^{12}$, there are a few islands
as in Fig. 5(c), and $g(\bm r)$ has a broad peak at around $\bm r=(150, 200)$.

So far, 
we discussed an extreme case with the step separation $L_s=2$ in order to
realize a strong anisotropic effect in our small system. 
Since  the island shape anisotropy is expected to be caused by
an accumulation of many weak anisotropy effect at steps, one has to pack as many 
 steps in an island as possible. Actually, with $L_s=3$
we can observe similar arrangement of many anisotropic islands, but
with a larger separation $L_s \ge 4$, it becomes harder to observe
anisotropy in islands. In a real system of Pn molecules, step separations
are wider, 4.5nm to 10nm, but an island size is about 10$\mu$m and
it spans many steps. It may be interesting if there is a critical size
for an island to show a shape anisotropy.

\section{Summary and Discussions}

Motivated by heteroepitaxial growth experiments of long organic molecules
on a flat \cite{kambe+07} or on a vicinal surface
\cite{nishikata+07},
effect of detachment anisotropy on the island shape is studied
by means of kinetic Monte Carlo simulations of a lattice gas model.

Only with a molecular deposition 
followed by a surface diffusion, 
islands take a ramified dendritic shape, similar to DLA.
Addition of detachment from the edge
makes fat dendritic branches, 
and with a large enough detachment rate 
islands become compact in a square form.

If the long molecules are tilted in an epitaxial film,
the rate of detachment from island edges may depend on the
edge orientation. 
Simulation shows that the edge with a smaller rate of detachment
shows diffusional instability and forms a dendritic tip.
Resulting island shape looks similar to the one obtained in the experiment
\cite{kambe+07}.

On a vicinal surface, an interaction between an edge molecule
and a neighboring one
depends on edge position.
When an edge has just climbed up the substrate step, 
a molecular bonding is weak and an edge molecule may easily  detach
from the edge site. On the other hand, when an edge has just stepped down
 the substrate step, a bonding is strong and the edge molecule 
 is hard to detach. In our simulation we assume an extreme case
of no detachment at a lower edge to stress an anisotropy effect.
Simulations show clear anisotropy in adsorbed island shape as
the step density increases.

In addition to the shape anisotropy of a single island,
the orientation-dependent detachment rate may induce a
spatial correlation in the arrangement of islands with clear peaks 
in the spatial correlation function.
Since the random deposition process introduces shot noise
in the nucleation process, a fast surface diffusion seems
necessary to average out fluctuation and to induce a spatial order,
but then the island separation becomes large.
For further studies on spatial correlation, one needs a larger system size.

As for the reason of island shape anisotropy, one may think about a
diffusional anisotropy, such that a step provides an asymmetry
in an energy barrier for the admolecule diffusion.
However, the asymmetric potential is unable to lead unidirectional
motion nor a net drift, as the Feynman's ratchet and pawl machine cannot
provoke directional rotation
\cite{feynman+66}. Another effect that
steps may provide is a nucleation center by lowering the energy barrier.
But in the experiment there is no observation 
of alignment of nucleation centers along
the step direction. Therefore, the latter effect might be small.
Observation of admolecule motion on the substrate, if possible, should 
resolve these points clearly.

\acknowledgments
%\section*{Acknowledgment}
We acknowledge G. Sazaki and S. Nishikata, and T. Nakada and T. Kambe
for showing their experimental data prior to the publication. 
The work is supported by a Grant in Aid for Scientific Research 
from Japan Society of the Promotion of Science, No. 19540410. 
Y. S. is benefited from the interuniversity 
cooperative research program of the Institute for Materials Research, 
Tohoku University.

%%%%%%%%%%%%%%%%%%%%%%%%%%%%%%%%%%%%%%%%%%%%%%%%%%%%%%%5

\end{document}